\documentclass[10pt,twocolumn,letterpaper]{article}

\usepackage{cvpr}              
\usepackage{graphicx}
\usepackage{amsmath}
\usepackage{amssymb}
\usepackage{booktabs}
\usepackage{algorithm}
\usepackage{algpseudocode}

\usepackage[dvipsnames]{xcolor}

\definecolor{cvprblue}{rgb}{0.21,0.49,0.74}
\usepackage[pagebackref,breaklinks,colorlinks,citecolor=cvprblue]{hyperref}

\def\paperID{00031} 
\def\confName{CVPR}
\def\confYear{2024}

\title{DISC: Latent Diffusion Models with Self-Distillation from Separated Conditions for Prostate Cancer Grading}

\author{
\begin{tabular}{@{}c@{}}
Man M. Ho$^{1}$ \qquad 
Elham Ghelichkhan $^{1}$ \qquad 
Yosep Chong$^{2,4}$ \qquad 
Yufei Zhou$^{3}$  \qquad \\
Beatrice Knudsen$^{4,5}$ \qquad 
Tolga Tasdizen$^{1,6}$
\end{tabular}
\\
$^{1}$ Scientific Computing and Imaging Institute, University of Utah, Utah, USA \\
$^{2}$ The Catholic University of Korea College of Medicine, Seoul, Korea \\
$^{3}$ Case Western Reserve University, Ohio, USA\\
$^{4}$ Departmant of Pathology, University of Utah, Utah, USA \\
$^{5}$ Huntsman Cancer Institute, University of Utah Health, Utah, USA \\
$^{6}$ Department of Electrical and Computer Engineering, University of Utah, Utah, USA \\
}

\begin{document}
\maketitle

\begin{abstract}
Latent Diffusion Models (LDMs) can generate high-fidelity images from noise, offering a promising approach for augmenting histopathology images for training cancer grading models. While previous works successfully generated high-fidelity histopathology images using LDMs, the generation of image tiles to improve prostate cancer grading has not yet been explored. Additionally, LDMs face challenges in accurately generating admixtures of multiple cancer grades in a tile when conditioned by a tile mask. In this study, we train specific LDMs to generate synthetic tiles that contain multiple Gleason Grades (GGs) by leveraging pixel-wise annotations in input tiles. We introduce a novel framework named Self-Distillation from Separated Conditions (DISC) that generates GG patterns guided by GG masks. Finally, we deploy a training framework for pixel-level and slide-level prostate cancer grading, where synthetic tiles are effectively utilized to improve the cancer grading performance of existing models. As a result, this work surpasses previous works in two domains: 1) our LDMs enhanced with DISC produce more accurate tiles in terms of GG patterns, and 2) our training scheme, incorporating synthetic data, significantly improves the generalization of the baseline model for prostate cancer grading, particularly in challenging cases of rare GG5, demonstrating the potential of generative models to enhance cancer grading when data is limited.
\end{abstract}
\section{Introduction}

\begin{figure}
    \centering
    \includegraphics[width=\linewidth]{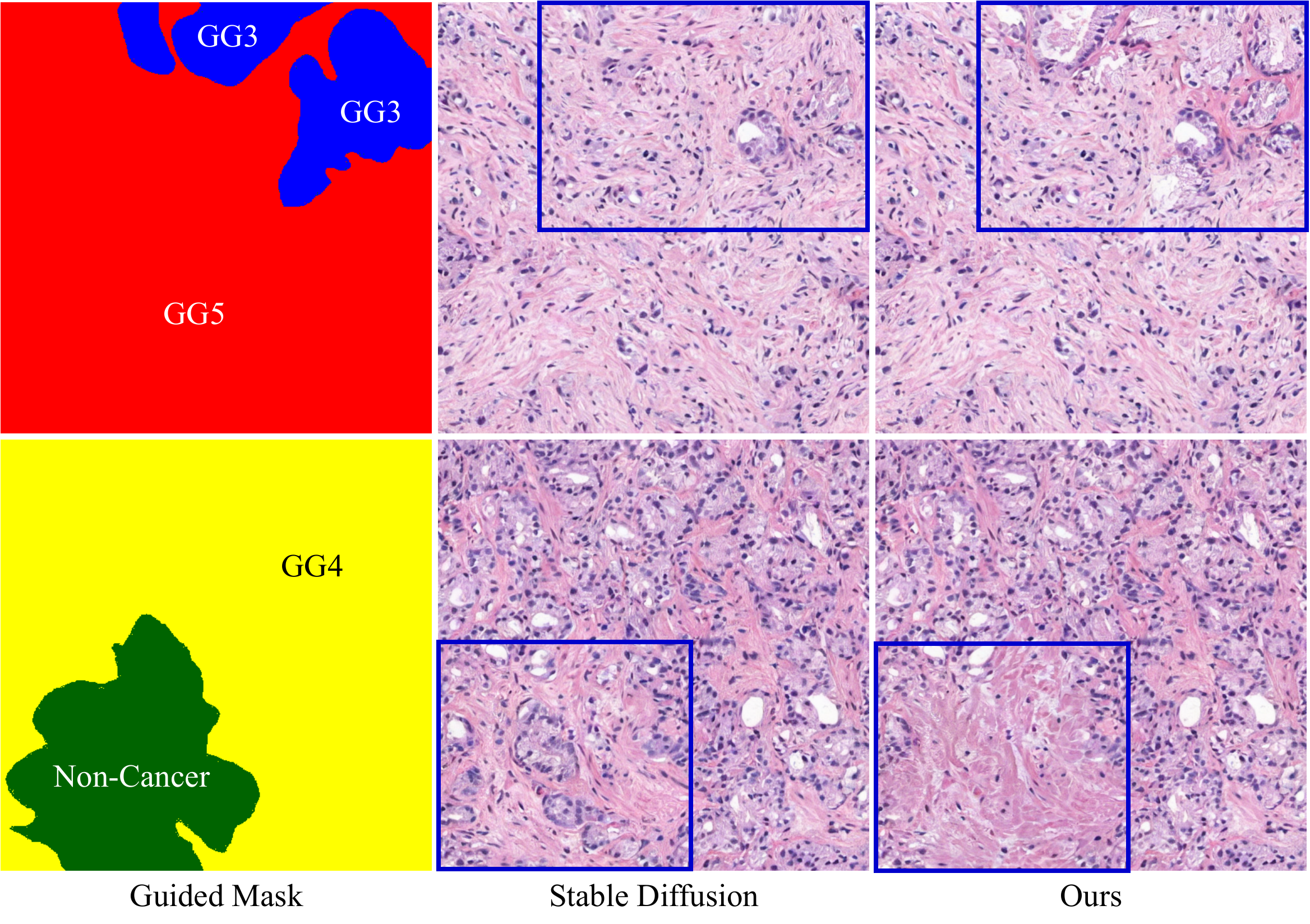}
    \caption{Stable Diffusion \cite{rombach2022high} produce a sheet of cells resembling GG5 in GG3-indicated regions (\textit{top}) and fused glands resembling GG4 in Non-Cancer-indicated regions (\textit{bottom}).}
    \label{fig:problem}
\end{figure}


\begin{figure*}
    \centering
    \includegraphics[width=0.8\textwidth]{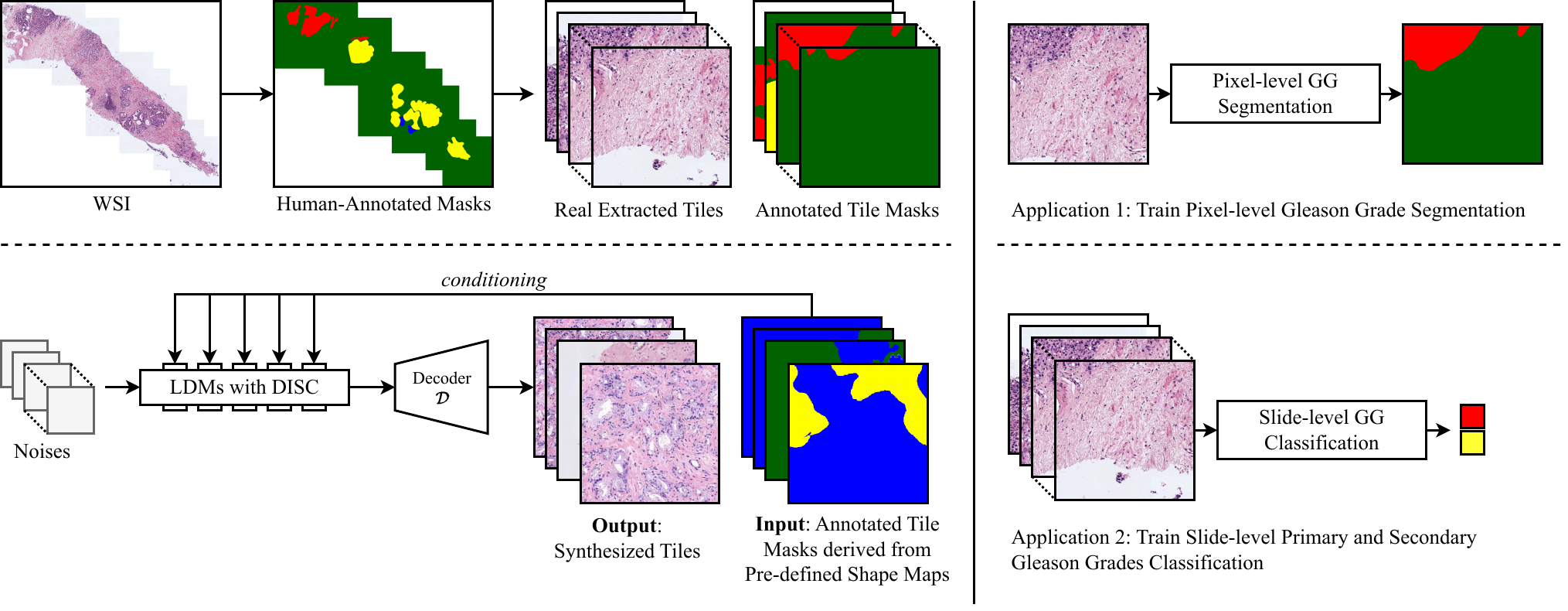}
    \caption{Besides the real patches (\textit{top-left}) for training pixel-level and slide-level Gleason grading models (\textit{right}), we introduce Latent Diffusion Models (LDMs) \cite{rombach2022high} with Self-\textbf{Di}stillation from \textbf{S}eparated \textbf{C}onditions (DISC) to accurately generate admixtures of multiple Gleason Grades in a tile when conditioned by a tile mask (\textit{bottom-left}).}
    \label{fig:overview}
\end{figure*}

In recent years, Latent Diffusion Models (LDMs) \cite{rombach2022high, moghadam2023morphology} have emerged as a powerful tool in computational pathology for generating high-fidelity tiles for Whole Slide Images (WSIs). Synthetic histopathology images potentially improve multiple downstream tasks, one of which is the training of cancer grading models (please refer to Supplementary Document for application overview). However, in prostate cancer grading, the current utilization of LDMs remains limited, as it does not effectively incorporate multiple cancer grades into synthetic tiles. In prostate cancer, the growth pattern of the cancer is used to define the cancer grade, which is named the Gleason grade. Pathologists identify 5 different Gleason Grade (GG) groups to forecast the severity of prostate cancer and likelihood of cancer progression. However, synthetic tiles generated by LDMs may display incorrect Gleason grade patterns, or mistake benign glands for high grade cancer (Fig. \ref{fig:problem}). To address these issues, we introduce a novel approach: we first tailor LDMs to produce tiles conditioned by human-annotated masks that feature multiple GG labels. Building on the principle of "\textit{Get More Done - One Thing at A Time}" \cite{zack2015singletasking}, we further refine this approach with our Self-Distillation from Separated Conditions (DISC) technique, aimed at improving the precision of GG patterns guided by intricate masks. Leveraging the methodology outlined in \cite{ho2022deep}, we also develop a training framework that efficiently utilizes generated tiles to enhance the performance of both pixel-level and slide-level cancer grading models, as illustrated in Figure \ref{fig:overview}. Moreover, we implement a straightforward yet effective sampling strategy to ensure a balanced representation of GGs within the tile masks, thus addressing potential label distribution imbalances in the training dataset. Our work is available at \url{https://minhmanho.github.io/disc/}.

\textbf{Advancements in Histopathology Image Synthesis}. Following the success of Generative Adversarial Networks (GANs) \cite{goodfellow2020gan1, karras2017gan2, brock2018gan3, kang2023gan4} in image synthesis, Diffusion Models have become a leading approach for generating high-fidelity images from noise \cite{sohl2015dm0, ho2020denoising, song2020denoising, saharia2022dm1, zheng2023layoutdiffusion}. Rombach et al. \cite{rombach2022high} have significantly advanced the field with the introduction of Latent Diffusion Models (LDMs), which demonstrate exceptional image synthesis capabilities with reduced computational demand by utilizing pre-trained autoencoder-based latent spaces. These innovative generative models \cite{moghadam2023morphology} are revolutionizing computational pathology by providing robust data augmentation capabilities for a variety of downstream applications, including nuclei segmentation \cite{hou2019nucleiseg, butte2023nucleiseg, ding2023large}, polyp segmentation \cite{thambawita2022singan}, the analysis of skin lesions, and the classification of Renal Cell Carcinoma (RCC) \cite{chen2021synthetic}. In our work, we specifically tailor LDMs to generate image tiles guided by complex masks that incorporate multiple Gleason Grades (GGs). Furthermore, we introduce Self-Distillation from Separated Conditions (DISC), an innovative method aimed at improving the precision of label patterns in the guided mask. Through the training of pixel-level and slide-level cancer grading models, such as Carcino-Net \cite{lokhande2020carcino} and TransMIL \cite{shao2021transmil}, alongside our synthetic tiles, we observe significant performance improvements, especially in diagnosing rare cases like GG5.

\textbf{Knowledge Distillation (KD) for Generative Models}. 
KD is a technique that transfers knowledge from a larger, more complex model (teacher) to a smaller, simpler model (student) \cite{hinton2015distilling}. In image classification, self-distillation, introduced in \cite{zhang2019your}, distills knowledge from deeper classifiers to shallower ones in neural networks. Self-distillation with no labels (DINO) \cite{caron2021dino} employs co-distillation \cite{anil2018large} to enhance the performance of Vision Transformers \cite{dosovitskiy2020image}. KD is also used in GAN-based image synthesis to improve results and computational efficiency \cite{wang2018kdgan, yuan2019ckd, li2020semantic}. For example, Self-distilled StyleGAN \cite{mokady2022self} filters uncurated internet images using a pre-trained StyleGAN and fine-tunes the model to generate images closer to cluster centers defined by the latent space. Meanwhile, KD has been applied for LDMs to improve sampling efficiency \cite{salimans2022progressive, meng2023distillation}. Inspired by \cite{zack2015singletasking}, we then separate the mask into single label masks and denoise latent features with one mask at a time, resulting in higher-confidence patterns for labels indicated in the label-guided mask. Finally, we propose Self-Distillation from Separated Conditions (DISC) to optimize computational cost while improving the quality of generated patterns.

Our contributions are as follows:
1) We propose the application of Latent Diffusion Models (LDMs) to generate histopathology patches using guided masks with multiple Gleason Grades.
2) We address the issue of LDMs generating incorrect labels when complex masks are provided by introducing Self-Distillation from Separated Conditions (DISC).
3) Our work surpasses previous studies in two aspects: (a) LDMs with DISC produce more accurate histopathology images compared to LDMs \cite{rombach2022high}. (b) Training baseline models such as Carcino-Net \cite{lokhande2020carcino} and TransMIL \cite{shao2021transmil} with our generated tiles leads to significant improvements on both in-distribution SICAPv2 \cite{silva2020going} and out-of-distribution LAPC \cite{li2018lapc} and PANDA \cite{bulten2022panda} datasets, particularly for the rare case of Gleason Grade 5 with limited data. This highlights the potential of generative models in enhancing rare cancer grading/detection with limited data.

\section{LDMs with DISC for Cancer Grading}
Latent Diffusion Models (LDMs) have shown their capability of generating high-fidelity images from noises, creating a promising approach for augmenting histopathology images in training cancer grading models. Although the previous works can generate high-fidelity histopathology images using LDMs, generating histopathology images with multiple Gleason Grades (GGs) is not entirely exploited, and the utilization of these generated images to improve the downstream task like pixel-level and slide-level Prostate Cancer (PCa) Grading is still an open question. Besides, LDMs still suffer from generating histopathology images conditioned by complex masks, as shown in Figure \ref{fig:problem}. In this work, we present specific LDMs, which can generate multiple GGs by leveraging pixel-wise annotation masks, discussed in Section \ref{sec:ldms}. 
For slide-level cancer grading models that require training pairs as \{multiple tiles, primary and secondary GGs\}, we employ an efficient sampling technique. This strategy enables the generation of tile sets tailored to specific primary and secondary Gleason grades, as detailed in Section \ref{sec:sampling}.
Additionally, to address the limitations of LDMs, we introduce the Self-Distillation from Separated Conditions (DISC) method, aimed at producing more precise GG patterns in alignment with GG-guided masks, as explored in Section \ref{sec:disc}. 
Lastly, we design a training framework that employs the generated tiles to significantly improve the accuracy of existing pixel-level and slide-level cancer grading models, as detailed in Section \ref{sec:training}. The comprehensive process is illustrated in Figure \ref{fig:overview}.

\subsection{LDMs conditioned by Gleason Grades}
\label{sec:ldms}

\begin{figure}
    \centering
    \includegraphics[width=\linewidth]{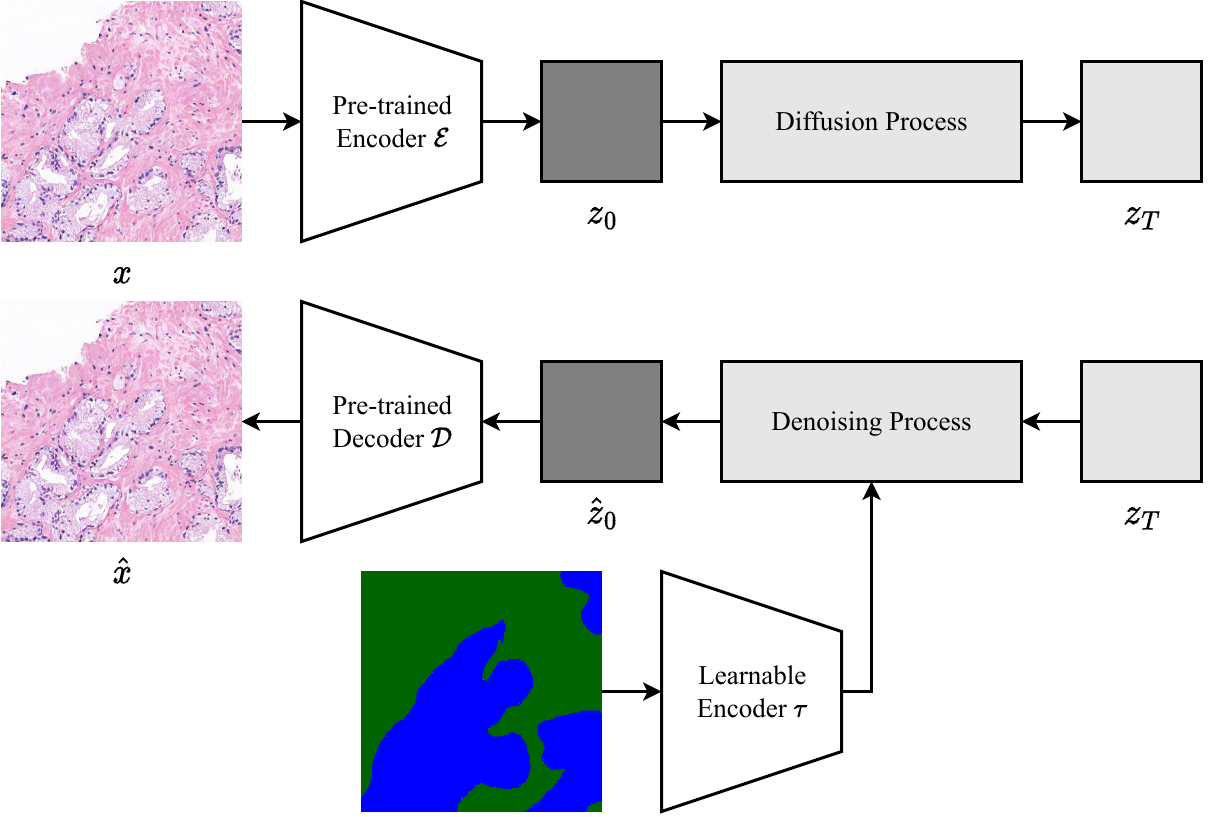}
    \caption{Latent Diffusion Models \cite{rombach2022high} conditioned by guided masks with multiple Gleason Grades (GGs)}
    \label{fig:sd}
\end{figure}

In this study, we train a generative model based on Latent Diffusion Models (LDMs) \cite{rombach2022high} (as known as Stable Diffusion) that is conditioned by guided masks with multiple Gleason Grades (GGs). Specifically, we consider an image tile $x \in \mathbb{R}^{H \times W \times 3}$ along with its pixel-wise annotated mask $m \in \{0, 1, 2, 3\}^{H \times W}$, where the labels $0, 1, 2, 3$ represent for Non-Cancer, GG3, GG4, and GG5, respectively. To extract essential features and reduce noises for high-quality image synthesis, we utilize a pre-trained VQ-regularization auto-encoder \cite{van2017neural} with encoder $\mathcal{E}$ and decoder $\mathcal{D}$ provided by \cite{rombach2022high} to encode and downsample the input image $x$ into a latent representation $z = \mathcal{E}(x)$, that $z \in \mathbb{R}^{h \times w \times c}$, with a factor $f = H/h = W/w = 4$. Subsequently, the decoder $D$ reconstructs the latent $z$ back to the input image $\hat{x} = \mathcal{D}(z) = \mathcal{D}(\mathcal{E}(x))$, as shown in Figure \ref{fig:sd}. In line with \cite{rombach2022high}, we employ a denoising U-Net \cite{ronneberger2015u}, denoted as $\epsilon_{\theta}$, to estimate the Gaussian noise $\epsilon \sim \mathcal{N}(0, I)$. This denoising model is conditioned by a GG-guided mask $m$, which is pre-processed by convolutional-layer-based $\tau_\theta$. Concretely, $m$ is fed to the encoder and decoder layers via cross-attention layers \cite{vaswani2017crossattn, jaegle2021crossattn, jaegle2021crossattn2, rombach2022high}. 
The objective is to minimize the simplified loss function:
\begin{equation}
   \mathcal{L}_{LDM} = \mathbb{E}_{z_0,m,\epsilon \sim \mathcal{N}(0, I),t}[||\epsilon-\epsilon_{\theta}(z_t, t, \tau_{\theta}(m))||^2_2]. 
\end{equation}
Here, $\epsilon_{\theta}$ and $\tau_\theta$ are jointly optimized, and $z_t$ represents the noisy version of $z_0$ at time step $t$, sampled uniformly from $\{1,...,T\}$, where $T=1000$. 

\subsection{Tile Annotation Mask Sampling}
\label{sec:sampling}

After training the denoising U-Net to predict added Gaussian noise accurately, we employ the DDIM sampler \cite{rombach2022high, song2020denoising} for faster sampling of image tiles conditioned by multiple Gleason Grades (GGs) with $T_{DDIM}=200$. To obtain and augment the annotation shapes, we preprocess existing human-annotated masks $m$ from SICAPv2 \cite{silva2020going}, converting them into tile shape masks, denoted as $m_{freq} \in \{0, 1, 2, 3\}^{H \times W}$. Here, labels are reclassified according to their frequency distribution, from the most to the least frequent (0-to-3). This approach aims to preserve the annotators' drawings, thus generating more authentic-looking tiles at low cost. 
In sampling phase, a tile shape map is randomly selected and non-overlapping cancer grading labels are assigned based on the random weights, which determines the label majority for a large number of $m_{freq}$. In simulating a tile set for slide-level classification via Multiple Instance Learning (MIL), we randomly select 20-100 annotation shape masks per designated primary and secondary GGs, with non-overlapping cancer grading labels distributed according to random weights. When the primary GG is Non-Cancer, the Non-Cancer label is exclusively applied. A comprehensive explanation and ablation study on Random Weights are in Supplemental Document.

\subsection{Self-Distillation from Separated Conditions}
\label{sec:disc}

While Latent Diffusion Models (LDMs) \cite{rombach2022high} are capable of producing high-fidelity tiles specifically designed for particular primary and secondary Gleason Grades (GGs) post-training, they still encounter difficulties in accurately generating Gleason patterns with high confidence for designated areas within the GG-guided mask. For example, when conditioned on pixel-wise human-annotated masks, LDMs might inaccurately generate a sheet of cells representing GG5 patterns in areas marked for GG3, where glandular structures are expected. Also, the Non-Cancer pattern generated by LDMs exhibits fused glands representing GG4 instead of stroma or uniform glands, as shown in Figure \ref{fig:problem}.

To address these issues, we draw inspiration from the characteristic of LDMs, which can generate high-confidence patterns for a single GG throughout the entire denoising process. We propose a denoising process with Separated Conditions (SC) for LDMs. At the start of the denoising process, we duplicate a Gaussian noise $z_T \sim \mathcal{N}(0, I)$ $K$ times to obtain a collection $\{z_T^0, \dots, z_T^{K-1}\}$, where $K=4$ represents the number of labels. Subsequently, LDMs denoise and infer $\{z_0^0, \dots, z_0^{K-1}\}$ at time step $t=0$ from $\{z_T^0, \dots, z_T^{K-1}\}$ conditioned by the corresponding separated masks ${sm_0, \dots, sm_{K-1}}$, where $sm_k \in \{k\}^{H \times W}$ and $k \in \mathbb{Z}, \quad 0 \leq k < K$. This enables the generation of $z_0^{k}$ with the strong characteristic patterns of label $k$. To generate the final latent feature representing the guided complex mask $m$, we downsample $m$ using nearest-neighbor interpolation and separate it into binary masks ${m_0, \dots, m_{K-1}}$, where $m_k \in {0, 1}^{h \times w}$ denotes the regions corresponding to label $k$ in $m$. We then multiply the latent features $z_0^{k}$ with the binary masks $m_k$ and merge them together to generate the final latent representation $z_0^{mixed}$ as:

\begin{equation}
\label{eq:ztmixed}
    z_t^{mixed} = Fuse(z_t^k, m_k) = \sum_{k=0}^{K} z_t^k \cdot m_k
\end{equation}

\begin{figure}
    \centering
    \includegraphics[width=\linewidth]{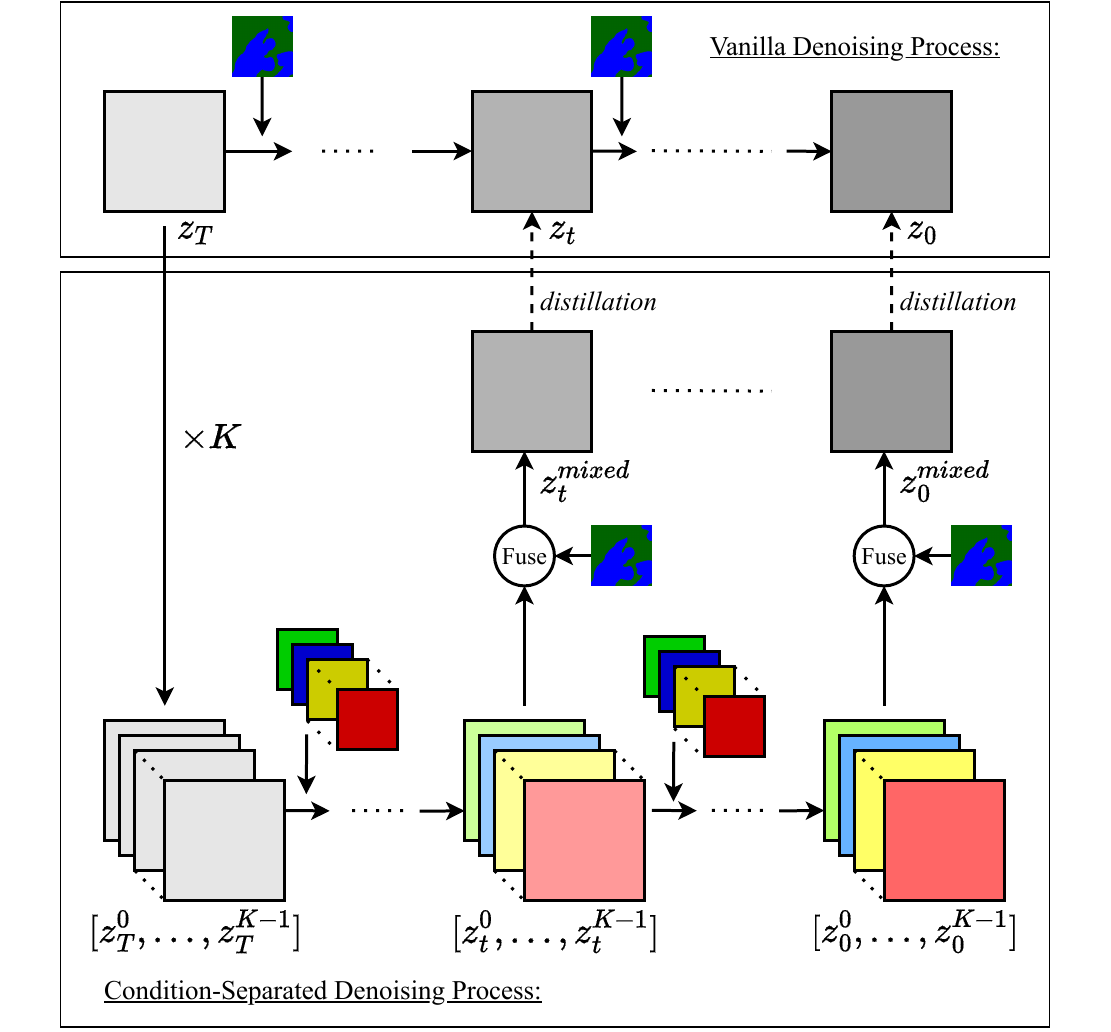}
    \caption{We introduce Self-\textbf{Di}stillation from \textbf{S}eparated \textbf{C}onditions (DISC) to improve image synthesis accuracy. Instead of using the initial complex guided mask with multiple Gleason Grades (GGs) (\textbf{top}), we generate separate latent features with distinct labels, which are fused with the mask in the final step for robust patterns. However, this approach incurs a computational cost of $\times K$, the number of labels. To address this, we train the main process to distill information from fused latent features obtained from the Condition-Separated Denoising Process (\textbf{bottom}).}
    \label{fig:disc}
\end{figure}

While the denoising process with SC enhances GG patterns in the generated tiles, it also increases time complexity by a factor of $K$. To maintain the speed of the vanilla denoising process conditioned by a complex mask, we propose Self-Distillation from Separated Conditions (DISC), where vanilla denoising process can mimic latent features $z_t^{mixed}$ from the denoising process with SC by optimizing $||z_t^{mixed}-z_t||_1$, as illustrated in Figure \ref{fig:disc}. To improve the efficiency of fine-tuning LDMs with DISC, we retain only the final latent features $z_0^k$ and define a simplified loss:

\begin{equation}
\label{eq:lossdisc}
\mathcal{L}_{DISC} = \mathbb{E}_{z_0^{mixed},m,\epsilon \sim \mathcal{N}(0, I),t}[||\epsilon-\epsilon_{\theta}(z_t^{mixed}, t, \tau_{\theta}(m))||^2_2]
\end{equation}

where the noisy $z_t^{mixed}$ can be obtained using a cummulative noise scheduler $\overline{\alpha}$ \cite{ho2020denoising} as $z_t^{mixed} = \sqrt{\overline{\alpha}_t} z_0^{mixed} + \sqrt{1-\overline{\alpha}_t} \epsilon$. Here, the generation of $z_0^{mixed}$ from $z_0^k$ with any random mask $m$ is achieved through Equation \ref{eq:ztmixed}. Eventually, we provide four models for further evaluation: 1) \textbf{SD}: Latent Diffusion Models (LDMs) \cite{rombach2022high}, also known as Stable Diffusion (SD), for generating tiles from a WSI conditioned by guided masks (top of Figure \ref{fig:disc}), 2) \textbf{SD-SC}: The pre-trained SD generates tiles with Separated Conditions (bottom of Figure \ref{fig:disc}), 3) \textbf{SD-DISC}: We generate $20,000$ samples of separated $z_0^k$ and continue to fine-tune the pre-trained SD exclusively on these samples, optimizing the loss function $\mathcal{L}_{DISC}$ from Equation \ref{eq:lossdisc} (top+bottom of Figure \ref{fig:disc}), and 4) \textbf{SD-DISC-CoTrain}: We also train SD-DISC with real data. This involves averaging the training errors from both $\mathcal{L}_{LDM}$ and $\mathcal{L}_{DISC}$.

\subsection{Training Prostate Cancer Grading Models}
\label{sec:training}
We demonstrate the effectiveness of our scheme in improving pixel-level and slide-level cancer grading performance of existing models by training CarcinoNet \cite{lokhande2020carcino} and TransMIL \cite{shao2021transmil} on both real and synthesized tiles from the SICAPv2 dataset \cite{silva2020going}. In slide-level grading, which involves predicting primary and secondary Gleason Grades (GGs), we modify the last layer of TransMIL from multi-class classification with a Softmax function to multi-label classification with a Sigmoid function. The TransMIL model is trained by minimizing the following loss function: $\mathcal{L}_{slide} = y\log(\hat{y}) + (1 - y) \log(1 - \hat{y})$.
Here, $y$ denotes the ground-truth label, and $\hat{y}$ represents the predicted label. To analyze the impact of generated histopathology images on improving the slide-level cancer grading model, inspired by \cite{ho2022deep}, we set a Balance Weight $\lambda \in [0, 1]$ to balance training errors between real and synthesized samples:

\begin{equation}
    \mathcal{L}_{total} = (1-\lambda)\mathcal{L}_{real\_slide} + \lambda\mathcal{L}_{synthesized\_slide}
\end{equation}

A higher value of $\lambda$ indicates a greater emphasis on optimizing the model using synthesized samples.

\section{Experiments}
In this section, we discuss the training and assessment of Latent Diffusion Models (LDMs) \cite{rombach2022high} using our proposed Self-Distillation from Separated Conditions (DISC) technique. Subsequently, we perform an ablation study on LDM conditions, including tile-level and pixel-level labels (please refer to the Supplementary Document for layouts \cite{zheng2023layoutdiffusion}). To establish our approach's superiority, we present a qualitative comparison among various models: vanilla Stable Diffusion (SD) \cite{rombach2022high}, SD with Separated Conditions (SD-SC), SD fine-tuned with DISC using 20,000 generated separated samples (SD-DISC), and SD-DISC fine-tuned with actual tiles (SD-DISC-CoTrain), all outlined in Section \ref{sec:disc}. Notably, we focus on qualitative assessment instead of relying on quantitative evaluation metrics such as FID and Inception Score (IS), which may not capture generated images with incorrect patterns. Moreover, we highlight the effectiveness of our generated data in enhancing both pixel-level and slide-level prostate cancer grading performance.
To achieve this, we train and compare the baseline models to those that have been jointly trained with tiles generated by our ablation models. Specifically, we utilize CarcinoNet \cite{lokhande2020carcino} as our baseline for pixel-level classification. For slide-level classification, TransMIL \cite{shao2021transmil} serves as the baseline, and Mixed Supervision \cite{bian2022multiple} is used as a comparison model.
Given that pixel-level annotations can be imprecise and incomplete \cite{anklin2021learning}, we qualitatively assess the segmentation results presented in this paper. Additionally, we perform a quantitative evaluation of the precision of pixel-level classification models, as detailed in Supplementary Document. For slide-level cancer grading, the models are evaluated using the Area Under the Receiver Operating Characteristic Curve (AUCROC) for the multi-label classification task of prostate cancer grading. Furthermore, we investigate how synthesized histopathology images affect the models' generalization by adjusting the balance weight $\lambda$ within the range of $[0.0, 0.9]$. All experiments are conducted on an NVIDIA RTX A6000 GPU.

\textbf{Training Latent Diffusion Models}. Firstly, we train SD \cite{rombach2022high} on two folds of the SICAPv2 dataset \cite{silva2020going}, where each fold has approximately 96 WSIs (7500 tiles) and 28 (2500 tiles) for training and validation, respectively. Following the completion of this dual-fold training phase (spanning 7 days), we select the model with the lowest validation error across both folds and transition to training it on the entire training dataset. This extended training phase for the chosen model spans 50 epochs, ensuring optimal generalization. Once SD is proficiently trained to generate prostate tiles, we then prepare 20,000 samples $z_0^k$ from Gaussian noise for fine-tuning using the Self-Distillation from Separated Conditions (DISC) technique. From this point, there are two pathways for further fine-tuning SD: (1) Fine-tuning on the 20,000 generated samples using DISC, denoted as SD-DISC, and (2) Jointly fine-tuning on both real tiles and the 20,000 generated samples, denoted as SD-DISC-CoTrain. It is important to note that SD-SC does not require additional training, as it utilizes the already well-trained SD model to generate condition-separated latent features. These pathways cost approximately 2-3 days.

\begin{figure*}
    \centering
    \includegraphics[width=\textwidth]{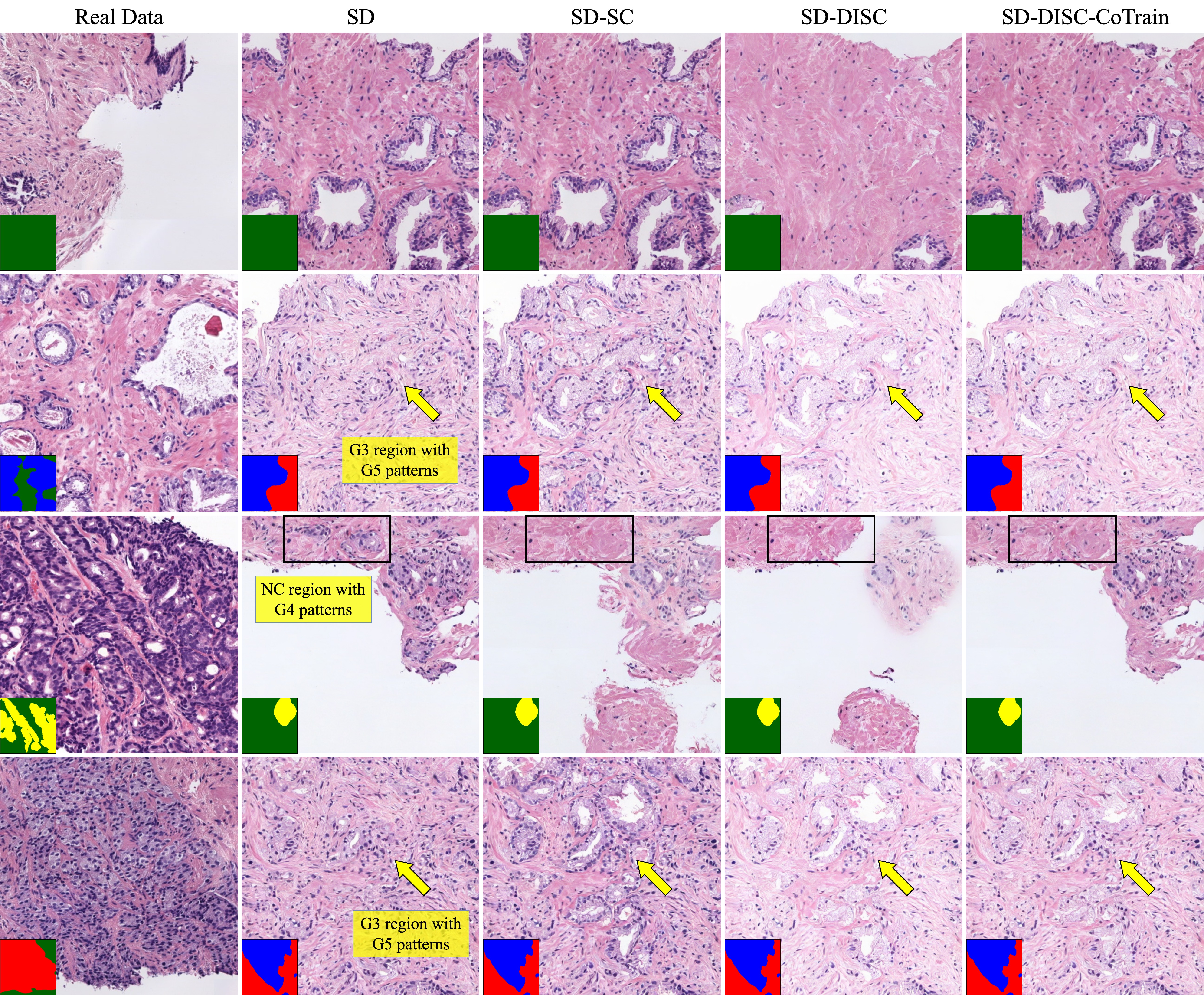}
    \caption{A qualitative comparison between Stable Diffusion (SD) \cite{rombach2022high} and our proposed technique, SD with Self-Distillation from Separated Conditions (DISC) (discussed in Section \ref{sec:disc}), for histopathology image synthesis. This work yields higher-confidence label patterns compared to SD. Notably, SD tends to generate fused glands representing GG4 for Non-Cancer regions (highlighted rectangles) and sheets of cells representing GG5 for GG3-indicated regions(\textcolor {yellow}{indicated by yellow arrows}). \textit{Labels}: \textcolor{green}{Non-Cancer}, \textcolor{blue}{GG3}, \textcolor{yellow}{GG4}, \textcolor{red}{GG5}.}
    \label{fig:qual_comparison}
\end{figure*}

\begin{figure*}
    \centering
    \includegraphics[width=\textwidth]{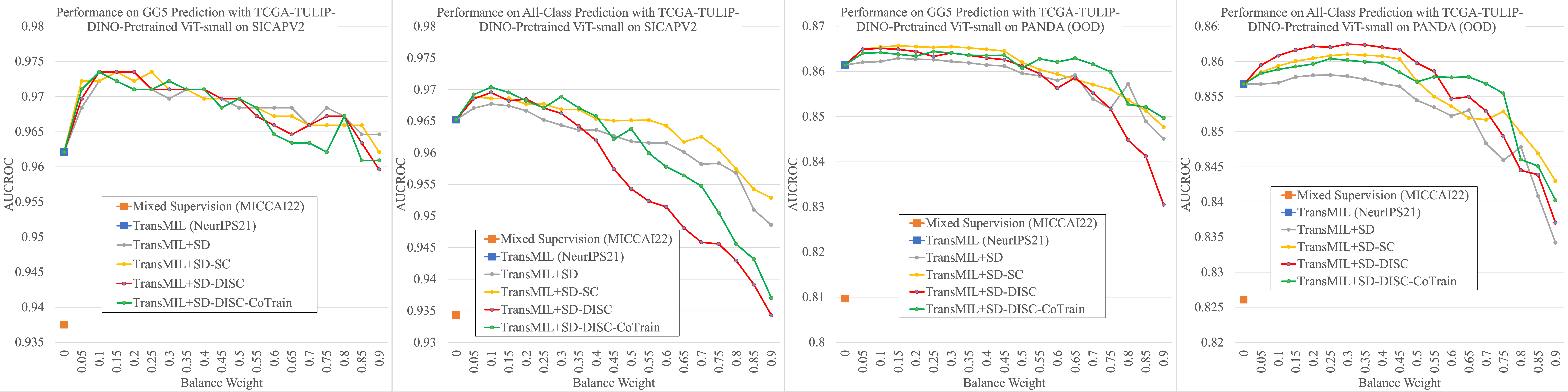}
    \caption{A quantitative comparison among TransMIL \cite{shao2021transmil}, Mixed Supervision \cite{bian2022multiple}, and TransMIL jointly trained with tiles generated by our models (discussed in Section \ref{sec:disc}) with a balance weight $\lambda \in [0.0, 0.9]$ in AUCROC. The feature representation extractor used is ViT-small (patch of 16) pre-trained on histopathology images with DINO \cite{caron2021dino, kang2023benchmarking}. All models are trained on the SICAPv2 \cite{silva2020going} and evaluated on both in-distribution SICAPv2 and Out-Of-Distribution (OOD) PANDA \cite{bulten2022panda}. Our generated data consistently improves cancer grading performance with higher AUCROC. Please check our Supplemental Document for more results including the feature representation extractors ResNet50 pre-trained on ImageNet and histopathology images with MoCov2 \cite{he2019moco, chen2020mocov2, kang2023benchmarking}.}
    \label{fig:quan_comparison}
\end{figure*}

\textbf{Training and evaluating cancer grading models}. 
We trained the TransMIL model \cite{shao2021transmil} and our ablation models using pre-extracted image tiles across 4 folds provided by the SICAPv2 dataset \cite{silva2020going}. Concurrently, Mixed Supervision \cite{bian2022multiple} employs a method of extracting tiles based on superpixel regions with centroid coordinates, similar to SegGINI \cite{anklin2021learning}. This strategy ensures more reliable instance-level labels, as patterns within the same region are more similar.
In our study, beyond utilizing existing tiles, we generated 276 tile sets representing 276 whole-slide images (WSIs) to balance the SICAPv2 dataset. The generation of a tile set, comprising 20-100 tiles, ranges from 4 to 16 minutes. The number of Whole Slide Images (WSIs) for each primary Gleason Grade (GG) was increased by generating additional tile sets as WSIs using models like SD, SD-SC, SD-SC-DISC, and SD-DISC-CoTrain, resulting in a complete set of 100 WSIs for every grade. We maintained consistency in tile generation for fair comparison by setting specific random seeds, which influenced the selection of GG-guided masks and Gaussian noise. For evaluation purposes, we not only utilized the test samples from the SICAPv2 dataset but also prepared a balanced test set with 100 WSIs for each label from out-of-distribution PANDA dataset \cite{bulten2022panda}. Additionally, we assessed the pixel-level performance of CarcinoNet using $2,200$ tiles from the LAPC dataset \cite{li2018lapc} focused on low-grade (GG3) and high-grade (GG4+GG5) cancer. To prepare the tiles for training and evaluation, we applied the tissue detection and tile extraction techniques described in CLAM \cite{lu2021data}, while Mixed Supervision relied on SegGINI for data preparation. For slide-level classification models \cite{shao2021transmil, bian2022multiple}, which depend on pre-trained embeddings, the extracted tiles are transformed into latent spaces using different models: ResNet50 pre-trained on ImageNet (a), ResNet50 pre-trained on TCGA and TULIP with MoCoV2 \cite{he2019moco, chen2020mocov2, kang2023benchmarking} (b), and ViT-small pre-trained on TCGA and TULIP with DINO \cite{caron2021dino, kang2023benchmarking} (c). 
The main paper includes results for (c), which demonstrated the most superior performance regarding feature representation and cancer grading compared to others. Please refer to the Supplemental Document for (a) and (b).

\begin{figure}
    \centering
    \includegraphics[width=\linewidth]{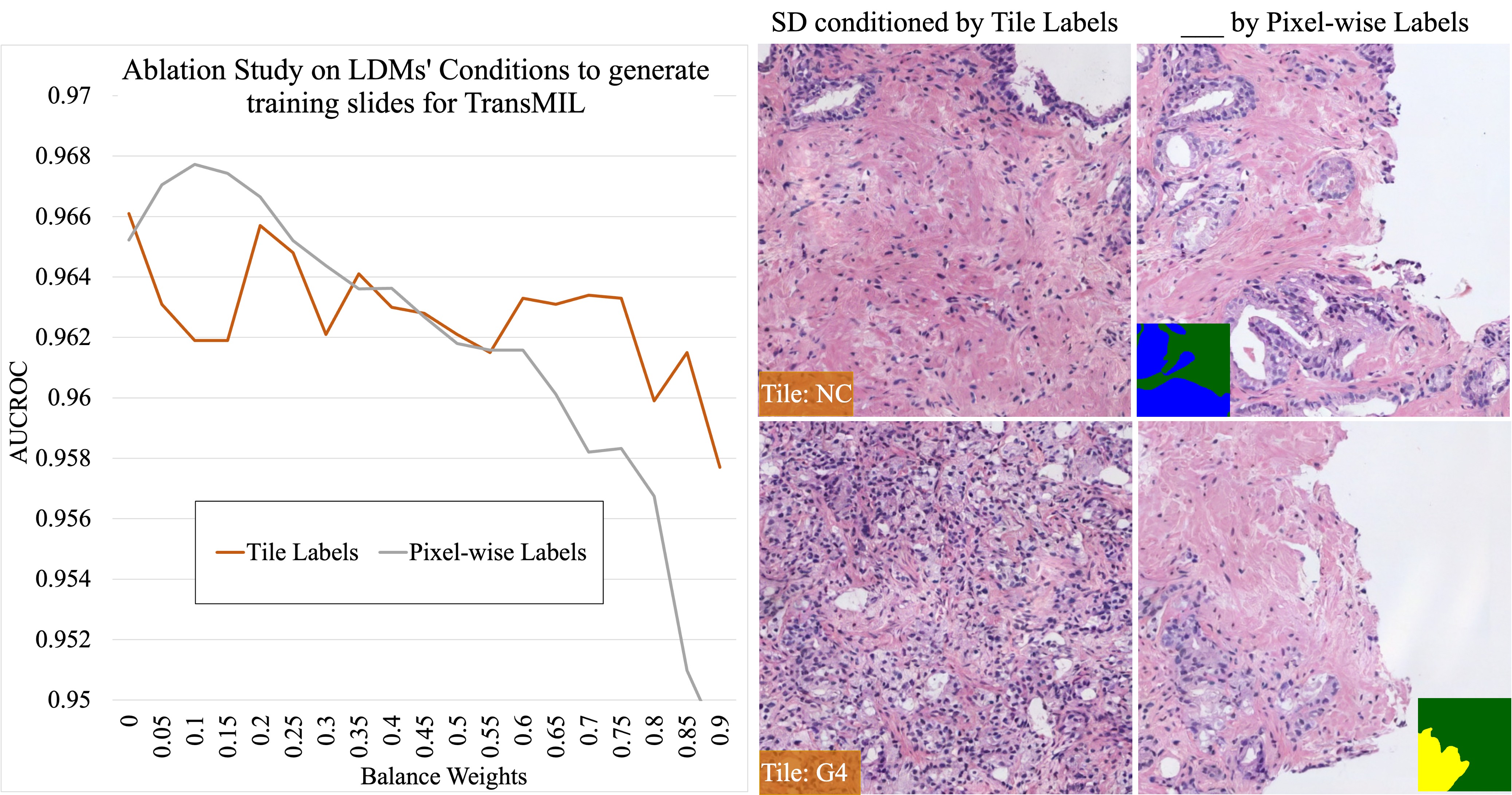}
    \caption{Ablation Study investigating the impact of Latent Diffusion Models' (LDMs) conditions including Tile Labels and our Pixel-wise Labels on enhancing TransMIL's performance (\textit{left}) in AUCROC and qualitative evaluation (\textit{right}). TransMIL utilizes feature representations pre-trained on histopathology images with ViT and DINO \cite{kang2023benchmarking}. Consequently, LDMs conditioned with Pixel-wise Labels effectively allow a mix of Gleason Grades (GGs) in the tiles. Conversely, LDMs conditioned with Tile Labels tend to generate a single pattern per tile. Quantitatively, TransMIL trained on tiles conditioned by Pixel-wise Labels achieves the best performance with a BW of 0.1.}
    \label{fig:ab_conditions}
\end{figure}

\begin{figure*}[t]
    \centering
    \includegraphics[width=\textwidth]{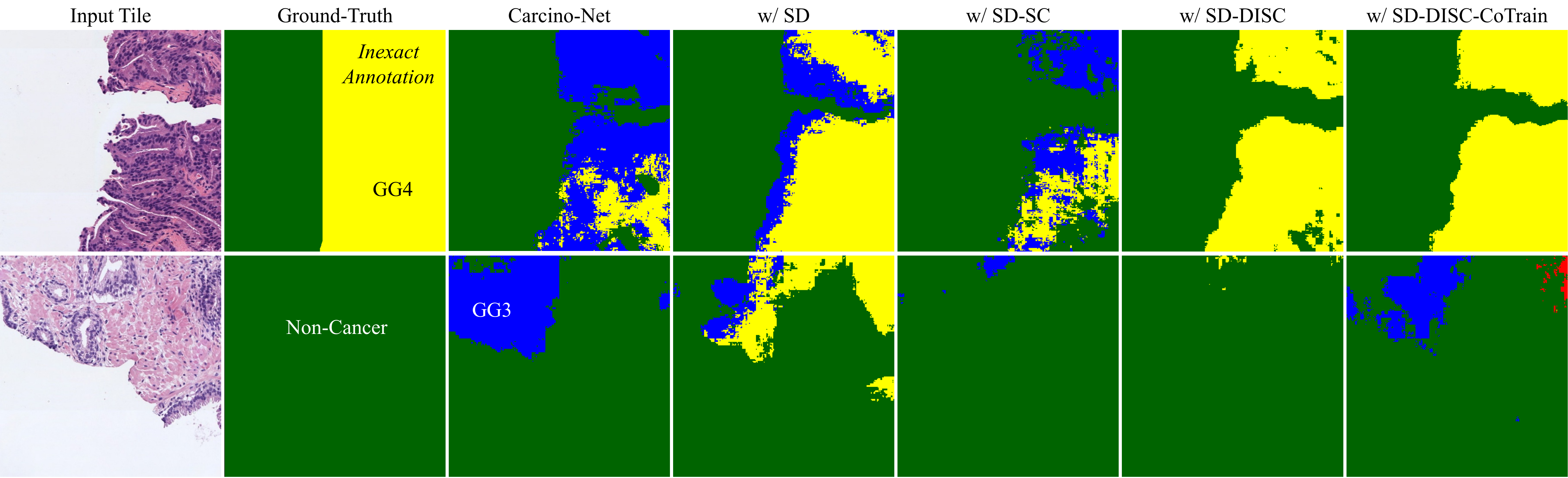}
    \caption{Qualitative comparison between Carcino-Net \cite{lokhande2020carcino} and itself trained with our techniques discussed in Section \ref{sec:disc}.}
    \label{fig:seg_qual_comparison}
\end{figure*}

\textbf{On LDM's conditions}.
Global labels, such as Tile Labels, provide weak information, causing Latent Diffusion Models (LDMs) to predominantly learn a standard pattern associated with the tile label while ignoring other patterns present within the training tile. Consequently, it becomes challenging for LDMs to generate admixture of Gleason Grades (GGs). Furthermore, combining Tile and Slide Labels is not a logically sound approach as they are independent variables for tile synthesis; however, we do present results from LDMs conditioned by this combination in Supplemental Document. To overcome this problem, we leverage pixel-wise labels available within SICAPv2 \cite{silva2020going} and propose an efficient sampling technique to automatically generate tile sets without requiring further user annotation. Consequently, tiles generated with pixel-wise labels contain the more anticipated patterns enriched with pixel-level information. Training TransMIL on such tiles yields the most optimal cancer grading performance, outperforming the two other conditions, as shown in Figure \ref{fig:ab_conditions}. Please refer to Supplementary Document for layouts as generation guidance.

\textbf{On improving the accuracy for histopathology image synthesis}.
In Figure \ref{fig:qual_comparison}, Latent Diffusion Models (LDMs) \cite{rombach2022high}, known as Stable Diffusion (SD), with pixel-wise labels successfully generate high-fidelity tiles (second column) that closely resemble the actual tiles (first column). Nevertheless, when conditioned by pixel-wise multiple-GG-guided masks, SD tends to generate incorrect patterns in certain regions. For instance, it fails to generate any glands in the GG3-indicated region (second and last rows) and produces fused glands representing GG4 in the Non-Cancer-indicated region (third row). To address these issues and enhance the accuracy of Non-Cancer and GG patterns, we introduce Separated Conditions (SC) to generate distinct latent features, denoted as SD-SC (third column). However, generating $K$ latent representations from $K$ label masks significantly increases the computational cost. To mitigate this challenge, we propose Self-Distillation from Separated Conditions (DISC) and fine-tune the well-trained SD using DISC, denoted as SD-DISC. Additionally, we train SD-DISC with real tiles to maintain realism, wherein training errors from real and synthesized tiles are averaged and jointly optimized, denoted as SD-DISC-CoTrain. As a result, SD-DISC can effectively mimic the latent features obtained from SD-SC, providing accurate patterns similar to SD-SC (fourth column). Nonetheless, these generated tiles occasionally deviate from realism, as observed in the third row. To address this limitation, we further train SD-DISC with real data, bringing the generated tiles closer to SD in terms of realism (last column). More results can be found in Supplementary Document.

\textbf{On improving pixel-level prostate cancer grading}. 
Our study aims to enhance the performance of pixel-level prostate cancer grading models by incorporating additional views of training tiles. To validate the effectiveness of our approach, we conduct both quantitative and qualitative comparisons with the baseline Carcino-Net \cite{lokhande2020carcino} and Carcino-Net trained on tiles generated by our ablation models (as detailed in Section \ref{sec:disc}). Our test sets comprise $2,100$ tiles from the in-distribution SICAPv2 dataset \cite{silva2020going} with Gleason Grade (GG) noisy pixel-wise annotations and $2,200$ tiles from the out-of-distribution LAPC dataset \cite{li2018lapc} for low-grade (GG3) and high-grade (GG4+GG5). The ground-truth annotations in the SICAPv2 dataset are inexact and incomplete, usually mislabeling Non-Cancer patterns such as background and stroma as GGs and providing incomplete annotations for GGs. These inaccuracies reduce the reliability of quantitative evaluations. Models trained on this imprecise ground-truth often struggle with misclassifying Non-Cancer patterns. Nevertheless, we still report on the cancer grading accuracy for positive predictions of Gleason Grades, omitting Non-Cancer label and focusing on pixel-level precision in Supplementary Document. As a result, Carcino-Net trained with SD-SC, SD-DISC, and SD-DISC-CoTrain effectively segments out Non-Cancer patterns thanks to generated training data with accurate annotations. In contrast, other models tend to misclassify Non-Cancer as Gleason patterns, as shown in Fig. \ref{fig:seg_qual_comparison}.

\textbf{On improving slide-level prostate cancer grading}. 
In this section, we assess the performance of previous works including TransMIL \cite{shao2021transmil}, its enhanced version Mixed Supervision \cite{bian2022multiple}, and TransMIL models jointly trained with tiles generated by our ablation models such as SD, SD-SC, SD-DISC, and SD-DISC-CoTrain. For all-class prediction, the baseline TransMIL model achieves an AUCROC of [\textbf{96.52\%}, \textbf{85.68\%}] on [\textbf{in-distribution SICAPv2}, \textbf{out-of-distribution PANDA}]. Meanwhile, tiles generated by SD yield improvements, resulting in [\textbf{96.77\%}, \textbf{85.80\%}] AUCROC ($\lambda$=0.1, $\lambda$=0.25). However, SD occasionally produces inaccurate patterns in specified regions, affecting the precision of synthetic training data. Addressing this, SD-SC is introduced and attains even better outcomes with AUCROC of [\textbf{96.89\%}, \textbf{86.01\%}] ($\lambda$=0.05, $\lambda$=0.3). SD-DISC-CoTrain, fine-tuned on in-distribution SICAPv2 while distilling from SD-SC, achieves top performance with an AUCROC of \textbf{97.04\%} ($\lambda$=0.1) on SICAPv2. SD-DISC provides more generalized training tiles for out-of-distribution PANDA, with TransMIL+SD-DISC achieving top performance with an AUCROC of \textbf{86.25\%} ($\lambda$=0.3). Additionally, our generative models improve performance for rare cases like GG5, with AUCROC improvements of up to [\textbf{+1.14\%}, \textbf{+0.57\%}]. Full and additional results on TransMIL enhancements with feature representations from two other extractors can be found in the Supplementary Document.

\section{Conclusion}
Latent Diffusion Models (LDMs) \cite{rombach2022high}, also known as Stable Diffusion (SD), have demonstrated their potential in augmenting histopathology image tiles for training cancer grading models. In this study, we trained LDMs conditioned by human-annotated masks with multiple Gleason Grades (GGs). Furthermore, we introduced SD with Separated Conditions (SD-SC), which generates distinct latent features conditioned by separated conditions, to enhance the accuracy of generating patterns indicated by the complex GG-guided masks. However, SD-SC is associated with an increase in processing time. To mitigate this computational cost while maintaining performance, we proposed SD with Self-Distillation from Separated Conditions (DISC), allowing the SD model to mimic the latent features of SD-SC and generate improved GG patterns. As a result, our LDMs with DISC can produce higher-confidence patterns for guided masks, as in Figure \ref{fig:qual_comparison}. Additionally, when using our augmented data, pixel-level and slide-level cancer grading models such as CarcinoNet \cite{lokhande2020carcino} and TransMIL \cite{shao2021transmil} demonstrate improved performance compared to their baselines, particularly in the challenging GG5 cases. Our approach also surpasses the advanced Mixed Supervision \cite{bian2022multiple} on both in-distribution and out-of-distribution data. In conclusion, our proposed LDMs with DISC offer a more accurate and effective approach for histopathology image augmentation, leading to improved cancer grading performance across different datasets and challenging GG categories.

\section{Acknowledgements}
We acknowledge the generous support from the Department of Defense Prostate Cancer Program Population Science Award (grant number W81XWH-21-0725); and also, the VA Merit Award (grant number 1 I01 CX002622-01). We also thank Dr. Akadiusz Gertych for the dataset from Cedars-Sinai Hospital in Los Angeles.

\section{Compliance with Ethical Standards}
LAPC images and annotations are available through a Material Transfer Agreement with Cedars-Sinai Hospital. The human subject data associated with the SICAPv2 dataset has been publicly released by \cite{silva2020going} under the Creative Commons Attribution 4.0 International license\footnote{\url{https://data.mendeley.com/datasets/9xxm58dvs3/1}}.

{
    \small
    \bibliographystyle{ieeenat_fullname}
    \bibliography{main}
}


\end{document}